\documentstyle[aps,epsf,preprint]{revtex}

\begin{document}
\draft
\preprint{HEP/123-qed}
\title{Effect of Chaotic Noise on Multistable Systems}
\author{Tsuyoshi Hondou}
\address{Yukawa Institute for Theoretical Physics \\
Kyoto University, Kyoto 606-01, Japan 
   }
\author{Yasuji Sawada}
\address{Research Institute of Electrical Communication \\ Tohoku University,
   Sendai, 980-77 Japan}
\maketitle
\begin{abstract}
 In a recent letter [Phys.Rev.Lett. {\bf 30}, 3269 (1995)],
we reported that a  
macroscopic chaotic determinism emerges in a multistable system:
 the unidirectional motion 
of a dissipative particle
subject to an apparently symmetric chaotic noise occurs
even if the particle is in a spatially symmetric potential. 
In this paper, we study the global dynamics of a dissipative particle
  by investigating 
the 
barrier crossing
probability of the particle
between two basins of the multistable potential.
We derive analytically an expression of the barrier crossing probability
of the particle subject to a chaotic noise generated by a general 
piecewise linear map.
We also show that the obtained analytical barrier crossing probability
is applicable to a chaotic noise generated  
not only by a 
piecewise linear map with a uniform invariant density but also
 by a non-piecewise linear map with non-uniform invariant density.
 We claim, from the viewpoint of the noise induced motion in a multistable
system, that chaotic noise is a first realization of the effect of 
{\em dynamical asymmetry} of general noise which induces the symmetry
breaking dynamics.
\end{abstract}

\section{Introduction}
 Chaotic systems show several unexpected and complex dynamics.
 "Chaotic itinerancy"\cite{Ikeda,Tsuda,Kaneko,Davis} and "evolution to edge of 
chaos"\cite{Packard,Langton,Kan,Bird}
are good 
examples. The mysterious role of chaos in neural networks has also been 
studied extensively\cite{Tsuda2,Freeman,Aihara,Inoue,Nara,Nozawa,Prog,Maru}.
 However, the origin of such interesting behaviors has not been clarified
 sufficiently;
 because an important feature of complex systems, multistability, 
has not been discussed explicitly in regard to the chaos. 

 The studies of multistable systems
subject to probabilistic noise
have extensively
been carried out in the field of reaction-rate theory, 
which is analyzed as stochastic processes\cite{RMP}.
The theory makes it possible
to calculate a barrier crossing probability in multistable
systems, in which the noise may have a simple time-correlation.
However, this theory also 
has a difficulty in treating a dynamical noise (perturbation),
especially for chaotic noise; because the theory
 is based on stochastic processes,
in which simple structure of the time-correlation of the noise is necessary
for its integrability.

 In addition to these background,
 some chaotic time series have been assumed to be
too random to retrieve its deterministic nature in physical systems, because
they may have the same randomness even as the coin tosses\cite{PhysicsToday}.
 Therefore the effect of chaotic noise  
has not been recognized as an important property even for a macroscopic
physical systems, whereas the chaotic noise has 
dynamical asymmetry\cite{def}.

 In the recent letter\cite{PRL}, we reported that the 
short-time correlation of chaotic noise caused by its determinism
is unexpectedly important in understanding the dynamics of multistable 
systems with
chaotic structures\cite{Chate}. 
 In this paper, we detail the analytical
 derivation of the barrier crossing probability
 of the dissipative particle in 
 multistable systems subject to chaotic noise and show
 that the analytical result is applicable to wider classes of chaotic noise.

 We also emphasize in this paper that chaotic noise is a first realization 
of the effect of dynamical asymmetry of any noise which induces unidirectional
motion of a dissipative particle in a {\em symmetric} potential. This is a 
new insight in regard to the discussions on the possible mechanism of protein motors
 by ratchet models\cite{Ratchet,Oosawa,Recent,temporal}.

In Section.II, we describe the system in which we will discuss the
effect of chaotic noise on the dissipative particle in a periodic
potential.
 In section.III we will derive the barrier
crossing probability of the dissipative particle over potential barrier,
where we use two kinds of generalized chaotic maps for wider application. 
In Section IV, we show that the present analytical result  
is applicable both to the chaotic noise which is non-piecewise linear map
and to the chaotic noise which
 has non-uniform invariant density, by using a logistic map chaos,
as an example.
In section.VI, we summarize our discussions,
where we remark the relation of our result to ratchet models
of protein motors.

\section{System}

 In this section, we describe a system 
 where we argue the effect of chaotic
noise on multistable systems.
 We discuss a dynamics of a dissipative particle in a periodic potential
subject to chaotic noise.
We believe that the present system is a minimal one 
which shows an effect of chaotic noise on a multistable system clearly.

 A dissipative particle in a potential $V$ and noise $\eta$ 
obeys the equation:
\begin{equation}
 \frac{dx}{dt}=-\frac{\partial V}{\partial x}+ \eta(t),
\label{eq:2}
\end{equation}
where  $\eta(t)$ is an additive noise.
We introduce a chaotic noise:
$\eta(t) = \sum_{j=-\infty}^{\infty} \eta_{j}
 \delta (t-j)$,
 where $\eta_{j}$ is a chaotic time series
 generated by:
\begin{equation}
 \eta_{n+1}=f(\eta_{n}),
\end{equation}
where $f$ is a chaotic map.
 The potential, $V(x)$, is any periodic 
potential. 
 In this paper, we report mainly the 
results of our study using a piecewise linear 
potential with parity symmetry:
 $V(x) = h-(h/L) |x( \, {
\rm mod}(2L))-L| $ for $ x \ge 0 $, $ V(-x) \equiv
V(x) $, where  $L$ is a half width of the period of the potential and $h$ 
is a height of the potential barrier.
We consider a system which satisfies the following condition:
\begin{equation}
0 < h/L < | \eta |_{\mbox{max}} \ll L.
\label{ine}
\end{equation}
In this condition,  the dissipative particle can 
move against the gradient of the potential
in both direction and the particle staying near a bottom of the potential
needs to be driven by chaotic noise many times to cross the potential 
barrier.  
In the following we study a discretized equation,
\begin{equation}
  x_{n+1}=x_{n}-\frac{\partial V}{\partial x}|_{x=x_{n}}+\eta_{n}
 \, \, \, (n=0,
\, 1, \, 2, \ldots),
\label{eq:3}
\end{equation}
which is approximately
obtained by integrating Eq.(1) from $t_{n}$ to $t_{n+1} = t_{n}+1$.
The choice of the finite $\Delta t \equiv t_{n+1} - t_{n} (=1)$ 
does not alter the central
result as shown in the following.

 We show here that the present system is a sufficient one which exhibit an 
unexpected dynamics under chaotic noise.
 We also show that the central result is not altered if the potential 
is not piecewise linear.
For the purpose, although this paper is intended 
 to discuss the effect of a general chaotic noise,
we briefly summarize the qualitative result of the symmetry breaking dynamics
by using the tent map chaos\cite{Grossmann,Schuster},
 $\eta_{n+1} = f (\eta_{n}) = 1/2 - 2 | \eta_{n}|$.
The tent map chaos has a uniform invariant density with parity symmetry,
$\rho ( \eta ) =1$ [ for $-0.5 \le x < 0.5$, otherwise $\rho (\eta) =0$ ]
and $\delta$ correlated\cite{Gade}, these
 properties of which are the same as the
uniform random number, $r_{n}$, $|r_{n}| < 0.5$. The tent map chaos
is one of the most random chaotic sequences which have the same randomness
as the coin tosses. Therefore 
the macroscopically broken parity
dynamics induced by this apparently symmetric chaotic noise had never 
been realized explicitly before our discovery, to our knowledge.

As found  in the previous letter\cite{PRL}, 
the chaotic noise generated by this tent map 
 can induce a broken symmetry dynamics of a dissipative particle
even in a symmetric multistable
system.
 It is easily verified that the qualitative results, namely, the broken
symmetry dynamics,
are  the  
same both for a smooth periodic potential and for a piecewise linear 
potential (Fig.2). 
If we replace the piecewise linear potential, $V(x)$, with the sinusoidal
potential having the same amplitude and the period, $V_{s}(x) = 
\frac{h}{2} sin(2 \pi x / (2L) -\pi /2)$;
the direction in which the particle moves does not change.
The quantitative
 difference of the velocity
as shown in Figure 1 can be attributed to the difference of the
 absolute
maximum gradients of the potentials:
$| \frac{\partial V_{s}}{\partial x}|_{\mbox{max}} = \frac{\pi}{2} | \frac{
\partial V_{0}}{\partial x}|_{\mbox{max}}$.
It can also be verified that a choice of $\Delta t \equiv t_{n+1} - t_{n}$
does not essentially alter the time evolution of the
dissipative particle
(Fig.2).

\section{Analytical derivation of barrier crossing probability}
 In this section, we argue a  
barrier crossing probability of a particle
 in a periodic potential subject to chaotic noise\cite{Journal}.
An average velocity of the particle is expressed in terms of the 
barrier crossing probabilities:
\begin{equation}
\langle v \rangle = 
\lim_{n \rightarrow \infty} \frac{x_{n}-x_{0}}{n} 
=  2L \{ \sum_{i} p_{i}^{+} - \sum_{j} p_{j}^{-} \},
\end{equation}
where $p_{i}^{+}$ is a barrier crossing probability in a positive direction caused 
by a process, $i$, and $p_{i}^{-}$ is that in a negative direction.
As is found later, the average velocity is often dominated by one barrier  
crossing probability, p:
$| \langle v \rangle | \sim 2L \cdot p$.


 When the slope of the potential is large enough, a particle is found mostly
in the neighborhood of one of the basins of the potential.
 Therefore, the particle needs
to be forced continuously by the noise having the coherent values to cross 
the barrier. Chaotic noise works effectively for the barrier crossing when
the noise stays in the neighborhood of an unstable fixed point, $\eta^{\ast}$.
There are two types of the chaotic sequences staying near an unstable
fixed point: 
One is the chaotic sequence leaving the unstable fixed point 
monotonically in its stay and the other is the sequence leaving
it with oscillation.
First, we discuss the former case which is relatively simple to treat.

In this paper, we have restricted ourselves for the chaotic noise in which
the two successive events of clustering around an unstable fixed point is not 
strongly correlated; in other words, the successive clustering does
not occurs without a sufficient intermission. 
However, there exists a case where the two successive events 
can be strongly correlated.
Bernoulli shift chaos is the case\cite{PRL}.
We will not discuss the complex case in this paper. 
The study is under way.

\subsection{Chaotic sequence monotonically leaving an unstable fixed point}

As shown in Fig.(\ref{fig:1}), the nearer 
the injected chaotic noise $\eta$ 
is to the unstable fixed point $\eta^{\ast}$, the
longer $\eta$ stays in the neighborhood of $\eta^{\ast}$. Therefore, we have to
calculate first how near the chaotic noise needs to be injected in the 
neighborhood of an unstable fixed point for the particle to cross the 
barrier.

 In the following, we calculate the maximum distance $\Delta_{c}$ 
between an injected 
chaotic noise and an unstable fixed point, for the barrier
crossing. The maximum distance, $\Delta_{c}$,
 is necessary to obtain the 
barrier crossing 
probability for the particle under chaotic noise. 
 To make the following discussion applicable to wider classes
 of chaotic maps, 
 we investigate the effect of a chaotic noise generated by a generalized 
piecewise linear map, 
which is characterized by the absolute value of the slope of the map, $\Lambda$,
 and an unstable fixed point, $\eta^{\ast}$.


 In a system which satisfies Eq.(\ref{ine}), the particle is mostly found near one of the 
bottoms of the potential. Therefore
 we assume that the particle is at an origin of the 
$x$-coordinate at discrete time, $n=1$, namely, $x_{1}=0$, when the chaotic
noise starts to drive the particle to cross the potential barrier. 
We also 
assume that
 the particle crosses the barrier by N ($ \gg 1$) time steps. 
 The nearer the chaotic sequence, $\eta_{n}$, is injected to an unstable
fixed point, the longer the particle continues to climb the potential.
 If the particle moves over a half width of the potential, $L$, 
within $N$ time steps, we judge that the particle crosses the potential
barrier. In this consideration,
we calculate the maximum distance 
 $\Delta_{c}= | \eta^{\ast} - \eta_{1}|$ for the barrier crossing
due to the effect of an unstable fixed point,
 as a function of $\eta^{\ast}$, $\Lambda$, $h$ and $L$. 
 Note that the slope of the potential $|\Lambda| > 1$ because the fixed point
$\eta^{\ast}$ is unstable. 

 From the conditions, we write the following inequality in which
 the particle
can cross the barrier:
\begin{equation}
 \eta_{N} \ge h/L,
\label{eq:a1}
\end{equation} 
\begin{equation}
 x_{N+1} \ge L,
\label{eq:a2}
\end{equation}
where we assume that the sign of $\eta$ is plus for simplicity; but the 
following result is also valid for negative $\eta$ by replacing it with
 an absolute value
of $\eta$. 
In such a case, the direction of the velocity of the particle is reversed.

 Let us set 
\begin{equation}
\eta_{1} = \eta^{\ast} - \Delta \, \, \,  (\Delta \ll 1),
\label{eq:a3}
\end{equation}
then one finds,
\begin{equation}
\eta_{N} = \eta^{\ast} - \Lambda^{N-1} \Delta.
\label{eq:a4}
\end{equation}
 We obtain through Eq.(\ref{eq:3}): 
\begin{equation}
 x_{N+1}  = \sum_{k=1}^{N}\eta_{k} - \frac{h}{L} (N-1),
\label{eq:a5}
\end{equation}
 where we use an approximation that $x_{1}=0$. 
Inserting Eq.(\ref{eq:a4}) into Eq.(\ref{eq:a5}) and using Eq.(\ref{eq:a2}), one obtains
\begin{equation}
 x_{N+1}=N \eta^{\ast} - \frac{\Delta (1-\Lambda^{N})}{1-\Lambda}
- \frac{h}{L} (N-1) \ge L.
\label{eq:a6}
\end{equation}
 Inserting Eq.(\ref{eq:a4}) into Eq.(\ref{eq:a1}), we get
\begin{equation}
 \Delta \le (\eta^{\ast} - \frac{h}{L})/\Lambda^{N-1}.
 \label{eq:a7}
\end{equation}
 Eq.(\ref{eq:a7}) gives the maximum $N$ for which the noise keeps driving the particle upward against the potential as a function of $\eta^{\ast}$, $\Delta$
 and $h/L$,
\begin{equation}
 N \le N_{0} \equiv  \log_{\Lambda}[(\eta^{\ast} - \frac{h}{L})/\Delta]+1.
\label{eq:a8}
\end{equation}
Replacing $N$ of Eq.(\ref{eq:a6}) by $N_{0}$ and approximating $1-\Lambda^{N}$
of Eq.(\ref{eq:a6}) by $-\Lambda^{N}$ for $N \gg 1$, we obtain the maximum value of $\Delta$ for which Eq.(\ref{eq:a2}) is satisfied, that the particle crosses
 the peak and make transition:
\begin{equation}
\Delta \le (\eta^{\ast} - \frac{h}{L}) 
\Lambda^{-\frac{(\Lambda-1)L^{2}+L \eta^{\ast} - \Lambda h}{
(\Lambda -1)(L \eta^{\ast}-h)
}}  \equiv \Delta_{c}(L,h,\eta^{\ast},\Lambda).
\label{eq:lo1}
\end{equation}
This expression contains the previous result calculated only for the tent map
\cite{Journal}, which is directly shown by setting $\Lambda = 2$ and 
$\eta^{\ast} = 1/2$.

In the following, we clarify in what limit the simple expression of the 
maximum distance $\Delta_{c}$, which is comparable to the scaling of the 
barrier crossing probability P shown in the previous letter\cite{PRL},
 is obtained.
First, we rewrite the maximum distance, $\Delta_{c}$, 
obtained in Eq.(\ref{eq:lo1}):
\begin{equation}
 \Delta_{c} = (\eta^{\ast} - \frac{h}{L} ) \Lambda^{-\alpha},
\end{equation}
where 
\begin{equation}
\alpha =
 \frac{L(1+\frac{L \eta^{\ast} - \Lambda h}{(\Lambda - 1) L^{2}})}
{\eta^{\ast} - \frac{h}{L}}.
\end{equation}
By the following limits, we get:
\begin{equation} 
\alpha \, \, \, \, \, \stackrel{\frac{L \eta^{\ast} - \Lambda h}{(\Lambda - 1) L^{2}} \ll 1}
 {\longrightarrow} \, \,
  \frac{L}{\eta^{\ast} - h/L} 
\label{eq:f1}
\end{equation}

\begin{equation}
 \stackrel{\frac{h}{\eta^{\ast} L} \ll 1}
 {\longrightarrow}  \frac{L}{\eta^{\ast}}
\label{samePRL}
\end{equation}
 This yields the same scaling
 expression as the barrier crossing probability in the previous 
letter\cite{PRL}:
\begin{equation}
 P(L) \sim (1 / \Lambda)^{ L / |\eta^{\ast}|},
\label{eq:PRL}
\end{equation}
 if $\Delta_{c} \propto P$.
 The first limit in Eq.(\ref{eq:f1})
is valid if the width of the potential
barrier is sufficiently large. 
The second in Eq.(\ref{samePRL}) is valid if the the effect of the
potential gradient 
is sufficiently small. 

\subsection{Chaotic sequence leaving an unstable fixed point with oscillation}

In this subsection, we discuss the maximum distance, $\Delta_{c}$, for the case that
the chaotic noise leaves an unstable fixed point 
with oscillation (Fig.(\ref{fig:2})).
 In a similar way as Eq.(\ref{eq:a1}) and Eq.(\ref{eq:a2}), we write
the inequalities in which the particle can cross the potential barrier:
\begin{equation}
 | \eta_{N} - \eta^{\ast} | \le \eta^{\ast} -h/L,
\label{eq:b1}
\end{equation} 
\begin{equation}
 x_{N+1} \ge L.
\label{eq:b2}
\end{equation}
 The chaotic noise
works to drive the particle against the gradient of the potential
 up to $n=N$ if Eq.(\ref{eq:b1}) is satisfied.  
One finds that the noise may drive the particle
against the potential, even if Eq.(\ref{eq:b1}) is not satisfied:
 This happens when $\eta_{N} - \eta^{\ast} >
\eta^{\ast} - h/L$. However, this ambiguity in the inequality (Eq.(20))
 does not change the 
following result of $\Delta_{c}$
without a pre-factor as shown later
(Eq.(\ref{eq:bd1}) and Eq.(\ref{eq:bd2})).

 Let us set 
\begin{equation}
\eta_{1} = \eta^{\ast} \pm \Delta \, \, \, (\Delta \ll 1),
\label{eq:b3}
\end{equation}
then one finds,
\begin{equation}
\eta_{N} = \eta^{\ast} \pm (-1)^{N-1} \Lambda^{N-1} \Delta.
\label{eq:b4}
\end{equation}
 We obtain through Eq.(\ref{eq:3}): 
\begin{equation}
 x_{N+1}  = \sum_{k=1}^{N} \eta_{k} - \frac{h}{L} (N-1),
\label{eq:b5}
\end{equation}
 by setting $x_{1}=0$.
One obtains from Eq.(\ref{eq:b4})
\begin{equation}
\langle \sum_{k=1}^{N} \eta_{k} \rangle_{\pm} = N \eta^{\ast},
\label{eq:bb}
\end{equation}
where $\langle \, \,   \rangle_{\pm}$ means an average over the sign, $\pm$,
of $\eta_{1}$.
By use of this averaging procedure, Eq.(\ref{eq:b5}) is replaced by
\begin{equation}
 x_{N+1}  = N \eta^{\ast} - \frac{h}{L} (N-1).
\label{eq:bc1}
\end{equation} 
Inserting Eq.(\ref{eq:bc1}) into Eq.(\ref{eq:b2}), one obtains
\begin{equation}
 N \eta^{\ast} - \frac{h}{L} (N-1) \ge L.
\label{eq:b6}
\end{equation}
Inserting Eq.(\ref{eq:b4}) into Eq.(\ref{eq:b1}), we get 
\begin{equation}
 N \le N_{c} \equiv \log_{\Lambda} \frac{\eta^{\ast} - h/L}{\Delta} + 1.
 \label{eq:b7}
\end{equation}
 Eq.(\ref{eq:b7}) gives the maximum $N$ for which the noise keeps driving the particle upward against the potential as a function of $\eta^{\ast}$, 
$\Delta$ and $h/L$.
Replacing $N$ of Eq.(\ref{eq:b6}) by $N_{c}$, we obtain the maximum value of $\Delta$ for which Eq.(\ref{eq:b2}) is satisfied, that the particle crosses
 the peak and makes transition:
\begin{equation}
\Delta \le \Delta_{c} \equiv 
  (\eta^{\ast} - \frac{h}{L}) \Lambda^{- \frac{L-\eta^{\ast}}{\eta^{\ast}
 - h/L}},
\label{eq:e1}
\end{equation}
 where we used an approximation for $\Delta  \ll 1$.

By the following limits, we get simpler expressions of $\Delta_{c}$
like Eq.(\ref{eq:f1}) and Eq.(\ref{samePRL}):
\begin{equation}
 \Delta_{c} \propto   \Lambda^{- \frac{L-\eta^{\ast}}{\eta^{\ast} - h/L}}
 \, \, \stackrel{\frac{\eta^{\ast}}{L} \ll 1}
 {\longrightarrow} \,
 \Lambda^{- \frac{L}{\eta^{\ast} - h/L}}
  \, \,  \stackrel{\frac{h/L}{\eta^{\ast}} \ll 1}
 {\longrightarrow} \,
 \Lambda^{-\frac{L}{\eta^{\ast}}}.
\label{eq:f3}
\end{equation}
The last expression is the same as that obtained for the chaotic noise which
monotonically leaves  
 an unstable fixed point
 (Eq.(\ref{samePRL})). 

 We note that the main result is not altered even if {\em r.h.s.} of 
Eq.(\ref{eq:b1}) is replaced by an arbitrary constant, $A$: namely, 
\begin{equation}
 | \eta_{N} - \eta^{\ast} | \le A.
\label{eq:bd1}
\end{equation}
In this case, we obtain 
\begin{equation}
 \Delta \le \Delta_{c} \equiv A \cdot \Lambda^{-\frac{L-\eta^{\ast}}{\eta^{\ast}
- h/L}},
\label{eq:bd2}
\end{equation}
which is the same as the previous expression (Eq.(\ref{eq:e1})) without a 
pre-factor, $A$.

\subsection{Barrier crossing probability and $\Delta_{c}$}
In this subsection we argue the relationship between the maximum distance, $\Delta_{c}$,
  and the barrier crossing probability.
First, we restrict ourselves to the chaotic
noise generated by the tent map function
 for demonstration.
This function has two unstable fixed points,
$\eta^{\ast}_{-}=-1/2$ and $\eta^{\ast}_{+}=1/6$,
the effect of which corresponds to the case of Fig.(\ref{fig:1}) and
 Fig.(\ref{fig:2}) respectively.
The former case is discussed here (Fig.(\ref{fig:tenta})).
Suppose that the barrier crossing event of the particle starts when the chaotic
noise is injected in the $\Delta$-neighbor of the unstable fixed point,
$\eta^{\ast}$. Then the event occurs only 
when the chaotic noise was in an interval,
$\Delta / \Lambda$, just before the start of the barrier crossing
as shown in Fig.(\ref{fig:tenta}). 
Thus, the sum of the invariant density over the interval, $\Delta / \Lambda$, 
 gives the barrier crossing 
probability\cite{Note}.
Therefore, we get the barrier crossing probability of the particle
in a negative direction caused by the effect of the unstable
 fixed point, $\eta_{-}^{\ast}$,
of the tent map function:
\begin{equation}
 P_{-} = \Delta_{c}/ \Lambda,
\label{simp}
\end{equation}
because the invariant density of the tent map function is uniform:
$\rho (\eta) = 1$.

The barrier crossing probability in a positive direction caused by 
$\eta_{+}^{\ast}$ can be obtained in the same way (Fig.(\ref{inj2})).  
However, the barrier crossing
 probability in a positive direction is much smaller
than that in a negative direction, because the amplitude of the unstable
fixed point, $\eta_{+}^{\ast}$, in a positive direction is much smaller than
that in a negative direction: $| \eta^{\ast}_{-}| \gg | \eta^{\ast}_{+} | $.
This is immediately be confirmed by the expression:
$ P \propto \Delta_{c} \propto \Lambda^{-L/\eta^{\ast}}$ 
(see Eq.(\ref{samePRL}) and Eq.(\ref{eq:f3})).
Therefore, the effect of one unstable fixed point, $\eta^{\ast}_{-}$,
 dominates the overall barrier crossing 
probability of the system with a sufficient barrier width, $L$.

The comparison between the analytical result and the numerical one
is shown for the noise generated by the tent map chaos (Fig.\ref{Add}).
The present theory sufficiently predicts the exponential decrease rate  
of the barrier crossing probability as to the potential width, $L$, 
  for $L \gg |\eta|_{\mbox{max}}$. 
The disagreement of the constant factor 
especially for the case,
 $h/L = 0.2$, may be attributed to
the ambiguity
of the assumption of the initial condition of the particle:
 $x_{1}=0$. When the chaotic noise is generated by the tent map, the noise
drives the particle by $\eta_{0} \sim +0.5$ in the positive direction
 just before the coherent drives caused by $\eta^{\ast}_{-} = -0.5$
in a negative direction for the barrier crossing.
 The more the slope of the potential is, the stronger this effect
works, because this anti-drive effect due to $\eta_{0} \sim +0.5$ needs
additional kicks in a negative direction
 for the noise to drive the particle, roughly equal to 
$\frac{L}{\eta^{\ast} - h/L}$ as similar as in the next subsection.

In general, the invariant density of a chaotic map is not 
uniform\cite{Grossmann,Schuster}. Thus,
a simple analytical expression of the barrier crossing probability
 cannot generally be obtained. 
Therefore, we derive a formal solution of the barrier crossing probability.
Let a set $I$: $I = \{\eta| f( \eta ) \in U(\eta^{\ast};\Delta_{c})$ and
$\eta \not\in U(\eta^{\ast};\Delta_{c}) \}$,
where $U(\eta^{\ast};\Delta)$ is the $\Delta$ neighborhood of an unstable
fixed point, $\eta^{\ast}$, and $f$ is a chaotic map.
Then, the sum of the invariant density over a set $I$ gives the desired 
expression of the barrier crossing probability:
\begin{equation}
 P = \int_{\eta \in I} d\eta \, \rho(\eta),
\label{eq:lo2}
\end{equation}
where $\rho (\eta)$ is the invariant density of the chaotic map.

If the invariant density
of a chaotic map in the region, $I$,
 is so smooth that it can be
approximated by a constant in the small region,
 the scaling form of the barrier crossing probability, $P$, is the same
as that of $\Delta_{c}$ without a pre-factor (such as $\Lambda^{-1}$).
This explains the coincidence of the scaling forms between $\Delta_{c}$ 
(Eq.(\ref{samePRL})) and the barrier crossing
 probability, $P$ (Eq.(\ref{eq:PRL})).

\subsection{Intuitive interpretation of barrier crossing probability}

The analytical result of the barrier crossing probability is easily 
understandable by a physical insight.
As mentioned in the last subsection, 
the barrier crossing probability and the critical distance, $\Delta_{c}$,
can have the same
scaling form for several kinds of chaotic noise of which an invariant density 
 is so smooth that it can be approximated by a constant 
in a small region, $I$, for the injection to the unstable fixed point.
We discuss this case for simplicity.

 Then, the scaling form of the barrier crossing
 probability as in Eq.(\ref{eq:f1}) 
and in
Eq.(\ref{eq:f3}) is:
\begin{equation}
 P \propto \Delta_{c} \propto
  \Lambda^{- \frac{L}{\eta^{\ast} - h/L}}.
\end{equation}
 The factor, $\eta^{\ast} - h/L$, is a displacement of the particle
 when the particle is kicked 
by one chaotic noise at an unstable fixed point. Therefore the value,
$\frac{L}{\eta^{\ast} - h/L}$, gives the number of the kicks 
necessary for the particle 
at a basin
of the potential to cross the potential barrier; where we used an approximation
that a chaotic noise has almost the same value as the unstable fixed point, 
$\eta^{\ast}$, during the condition, Eq.(\ref{eq:a1}) or Eq.(\ref{eq:b1}),
is satisfied. The approximation can be justified at least for the piecewise
linear maps, as found in the analytical derivation of the
barrier crossing probability. 

Because the value, $\frac{L}{\eta^{\ast} - h/L}$, gives the number of the necessary
kicks by chaotic noise, we can understand the exponential dependence of the 
barrier crossing probability on the value, $\frac{L}{\eta^{\ast} - h/L}$.
Suppose that 
\begin{equation}
|\eta_{1} - \eta^{\ast}| \le \Delta_{s} 
\end{equation}
which satisfies Eq.(\ref{eq:a1}) or Eq.(\ref{eq:b1}) and 
the particle crosses the barrier.
If the factor, $\frac{L}{\eta^{\ast}-h/L}$, increases by one,
the following inequality for the noise, $\eta^{\prime}$,
  must be satisfied for the barrier crossing:
\begin{equation}
|\eta_{1}^{\prime} - \eta^{\ast}| \le \Delta_{s} \cdot \Lambda^{-1}, 
\label{eq:g}
\end{equation}
 because Eq.(\ref{eq:g}) is equivalent to
 $|\eta_{2}^{\prime} - \eta^{\ast}| \le \Delta_{s}$,
 where the noise continues to drive the particle  $N$
times after $n=2$.
 Thus the
 unit increase of $\frac{L}{\eta^{\ast}-h/L}$ decreases the measure of 
set $I$ by $\Lambda^{-1}$.
 Therefore, the barrier crossing probability depends exponentially on the value,
$\frac{L}{\eta^{\ast} - h/L}$, with the base, $\Lambda$, when the distribution 
of the invariant density is sufficiently smooth.

\section{An application to the logistic map}
In the previous discussions including Ref.\cite{PRL}, we 
have argued only the effect of 
 chaotic noise generated by a piecewise linear map with uniform
 invariant density. 
Therefore we consider in this section 
the validity of the present analytical results
by applying them to the map which is a non-piecewise linear map
and has a non-uniform invariant density.
For this purpose, we use a logistic map\cite{Grossmann,Schuster} 
as a chaotic noise,
because an analytical expression of the invariant density is available.
The chaotic sequence of the logistic map appears as follows:
\begin{equation}
 \eta_{n+1} = f (\eta_{n}) = 1/2 -4 \,  \eta_{n}^{2}.
\end{equation}
 The invariant density of the logistic map is,
 $\rho (\eta) = \frac{1}{\pi \sqrt{(\eta+1/2)(\eta - 1/2)}}$,
 $[ -1/2 \le \eta < 1/2, \mbox{otherwise}\,  \rho(\eta) = 0]$.

Because an analytical expressions of $\Delta_{c}$ for the barrier crossing 
probability were derived only for a piecewise linear map,
 we have to make an approximation 
 to obtain the barrier crossing probability induced by the logistic map.  
 As noted previously, 
 the clustering event of chaotic noise near an unstable fixed point
 occurs when $\eta_{n}$ is injected 
near to the unstable fixed point.
 Therefore, it seems valid to linearize the logistic map around an unstable
 fixed point ($\eta \sim -0.5$)
 and the corresponding region ($\eta \sim 0.5$) 
 to be injected
near to the unstable fixed point.
  With this procedure, we get the linearized slope
of the map: $\Lambda = 4$  (see Fig.\ref{fig:logi}).

 By use of this approximation, we can evaluate the barrier crossing probability for 
 a non-uniform invariant density by using Eq.(\ref{eq:lo2}).
 The set $I$ of Eq.(\ref{eq:lo2}) is: 
 $I = \{\eta| 1/2 - \Delta_{c}/4 \le \eta < 1/2\}$.
 Therefore, we get an analytical expression of the barrier crossing
 probability:
\begin{equation}
 P = \int_{1/2-\Delta_{c}/4}^{1/2} d\eta \frac{1}
{\pi \sqrt{(\eta+1/2)(\eta - 1/2)}}
 = \frac{2}{\pi} (\sin^{-1} \sqrt{\frac{\Delta_{c}}{4}}),
\label{theorlo}
\end{equation} 
where $\Delta_{c}$ is given by Eq.(\ref{eq:lo1}).

The theoretical barrier crossing probabilities 
of the logistic map are shown with the numerically
obtained barrier crossing 
probabilities (Fig.\ref{fig:logidata}).
The data of the larger slope of the potential ($h/L=0.2$) 
 agree better with the theoretical predictions (Eq.(\ref{theorlo}))
than that of $h/L=0.1$.
The disagreement for the smaller slope of the potential may be attributed
to the assumption of the present theory that
the particle should start climbing from $x_{1}=0$ when the chaotic sequence
is injected near to an unstable fixed point. If the slope of the potential
is small, the particle fluctuates much around one of the bottoms
 of the periodic potential. This fluctuation effect caused by
 the logistic map may be strong because the invariant density
goes to infinity at both edges of the map, $\eta = \pm 1/2$: this decreases the validity of the assumption 
of the present theory.

\section{Summary and discussion}
 We showed in this paper a new macroscopic feature of chaotic dynamics
 emerging in a multistable system:
the effect of chaotic noise on the multistable system is attributed
to its unstable fixed points, which reminds us of the deterministic
nature of chaos.
The new feature appears effectively
in a multistable system when the slope of the unstable
fixed point of the noise, $\Lambda$, 
is near the "edge of chaos", because the local 
Lyapunov index, $\lambda$, at the unstable fixed point is $\lambda
= \ln \Lambda$\cite{Note1}.
This is consistent at least with  recent literature of neural
 networks\cite{Maru,HondouPRE}
 
 The unidirectional motion of the dissipative particle in a periodic potential
has been discussed in relation to the dynamics of motor
 proteins\cite{Recent,temporal}. 
We showed that the unidirectional motion can be induced by an apparently
symmetric chaotic noise even if the particle is in a symmetric multistable
potential.
  Similar results have recently appeared with several 
variations\cite{temporal}
after our first report\cite{Journal}.
 The authors of the papers claim that the unidirectional
motion can occur in a symmetric multistable potential if an additive noise
is "temporally asymmetric." However, the asymmetric effect of the noises
they used can be attributed to asymmetric distribution of the probability
density of the noise. In this sense, the effect of the {\em 
temporally asymmetric} noise is rather static, and thus 
the effect of "temporally
asymmetry," should be discriminated from that of "dynamical asymmetry."  

Finally, we mention that the present analytical
 method to estimate barrier crossing probabilities 
 may be applied to the escape
rate problem induced by other types of time correlation of the 
fluctuation including
an intermittent chaos and non-chaotic time series:
 the escape rate is found to be strongly dependent on the {\em transient}
time-correlation of the additive noise\cite{Tasaki}.

\section*{Acknowledgement}

The authors would like to thank I.Nishikawa, T.Aoyagi, S.Sasa,  
K.Sugawara, K.Sekimoto, S.Tasaki, T.Chawanya, Y.Hayakawa and M.Sano 
for stimulating discussions and helpful comments.
This work is supported in part by the Japanese Grant-in-Aid
for Science Research Fund from the Ministry of Education, 
Science and Culture (No.
07740315).


\begin{figure}
\caption{Typical time evolution of the two kinds of systems
 under a tent map chaos: (1) a smooth periodic potential,
$V_{s}(x)=h/2 \sin(2 \pi x/(2L) - \pi/2)$,
and (2) the piecewise linear potential, 
where the same amplitude and the period of the potentials are used:
$L=5$ and $h=0.5$.}
\end{figure}
\begin{figure}
\caption{Dependence of a difference interval $\Delta t$ which is used to 
derive Eq.(4) from Eq.(1) on the evolution
of the system, where $L=5$ and $h=0.5$.
 Data for $\Delta t= 1$, $0.1$ and $0.01$ are shown. 
}
\label{fig:01}
\end{figure}
\begin{figure}
\caption{Emergence of short-time correlation (or clustering) 
of the chaotic noise
near an unstable fixed point (Case A): 
The chaotic sequence $\eta_{n}$ leaves an unstable fixed point 
$\eta^{\ast}$ of the piecewise linear map (solid line)
  monotonically and the
distance between 
$\eta^{\ast}$ and $\eta_{n}$ increases exponentially, where the dotted
line shows the line, $\eta_{n+1} = \eta_{n}$.
  The strong correlation starts when the noise is injected in
 the $\Delta$ neighborhood of the unstable fixed point.
 The less the slope of the map, $\Lambda$, the more the sequence stays
 near the unstable fixed point. 
}
\label{fig:1}
\end{figure}
\begin{figure}
\caption{Emergence of short-time correlation of the chaotic noise (Case B):
 The chaotic sequence $\eta_{n}$ leaves an unstable fixed point $\eta^{\ast}$
with oscillating around the unstable fixed point.
  The strong correlation starts when the noise is injected in
 the $\Delta$ neighbor of the unstable fixed point.
}
\label{fig:2}
\end{figure}
\begin{figure}
\caption{
Injection mechanism of chaotic sequence near to the $\Delta$-neighbor of the 
unstable fixed point, $\eta^{\ast}$ (Case A).
In this case of the tent map, the slope $\Lambda = 2$.
}
\label{fig:tenta}
\end{figure}
\begin{figure}
\caption{
Injection mechanism of chaotic sequence near to the $\Delta$-neighbor of the
unstable fixed point, $\eta^{\ast}$ (Case B).
In this case of the tent map, the slope $\Lambda = 2$.
\label{inj2}
}
\label{fig:tentb}
\end{figure}
\begin{figure}
\caption{Barrier crossing probabilities of the dissipative particle
subject to a tent map chaos, where the slopes of a piecewise linear potential
are $h/L = 0.1$ and $h/L = 0.2$. Both numerical and analytical results are 
shown.
}
\label{Add}
\end{figure}
\begin{figure}
\caption{
A form of a logistic map, $\eta_{n+1} = 0.5-4 \eta_{n}^{2}$.
Dotted lines show the linearized slope for approximation 
 of the analytical approach.
 The linearized slope, $\Lambda = 4$.
}
\label{fig:logi}
\end{figure}
\begin{figure}
\caption{Barrier crossing probabilities under the logistic map chaos,
where the slopes of the piecewise linear potentials are $h/L=0.2$ and
$h/L=0.1$.
Both numerical and theoretical results are shown.
}
\label{fig:logidata}
\end{figure}


\begin{thebibliography}{99}
\bibitem{def}
We use "dynamical asymmetry" to describe the property of any noise of which
the asymmetric effect is not attributed to its asymmetric 
probability distribution
but to its dynamics. The first usage
of the word appeared in T.Hondou and Y.Sawada, Report No.adap-org/9512002.
\bibitem{Note}
 The direct sum of an invariant density over
$\Delta$-neighborhood of an unstable fixed point
is not appropriate for the barrier crossing probability, because there is a
possibility of an overcounting of the probability
in one event for a barrier crossing.
The following sequence of the noise; $\eta_{1} = \eta^{\ast} +
\Delta_{c} /(2 \Lambda)$,
$\eta_{2} = f(\eta_{1}) = \eta^{\ast}+\Delta_{c}/2$, $\cdots$, for one barrier
crossing event gives an
example.
\end{thebibliography}
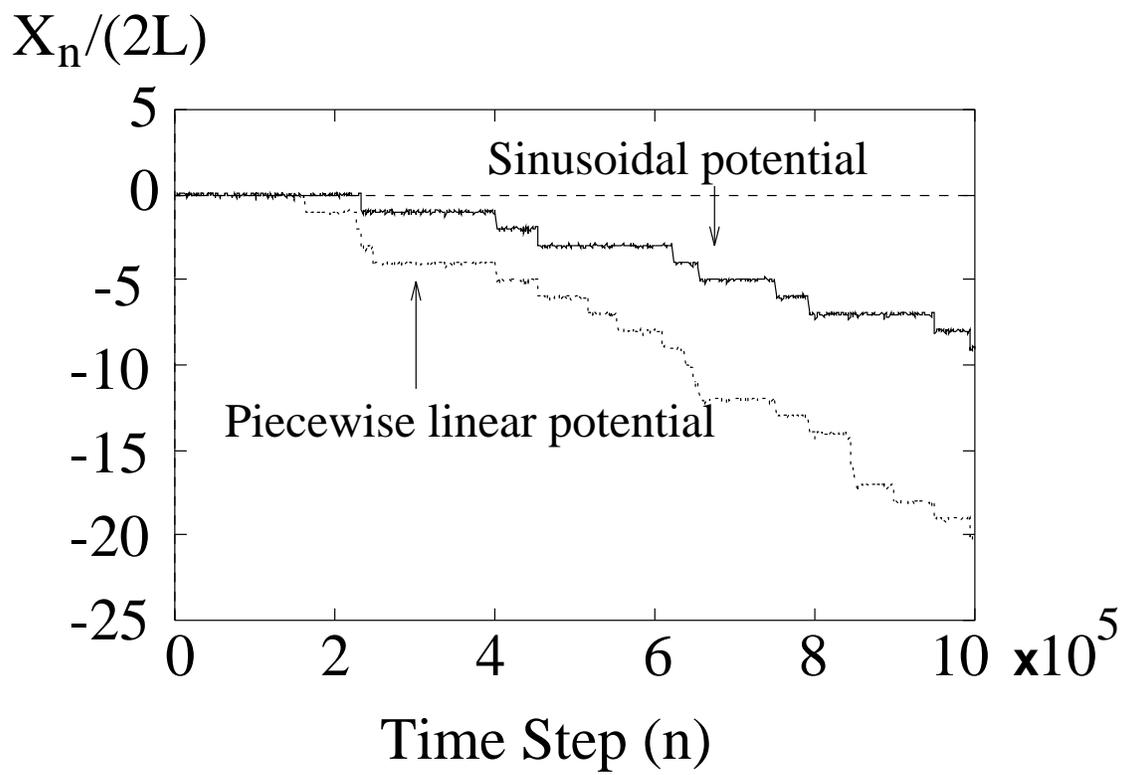
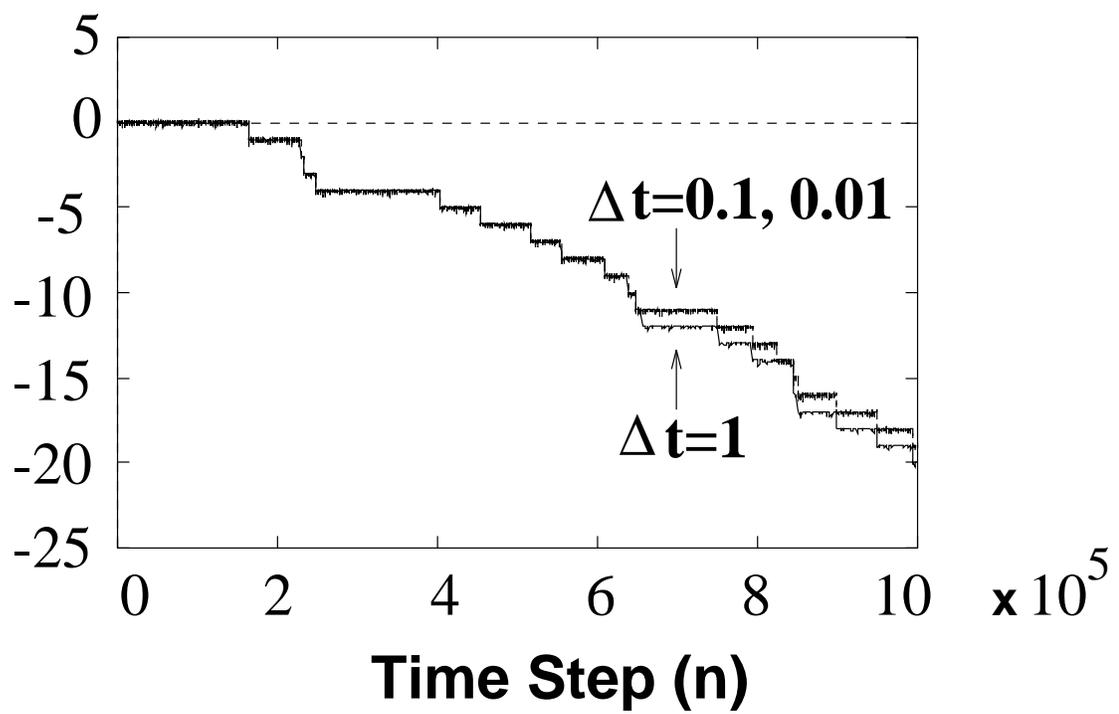
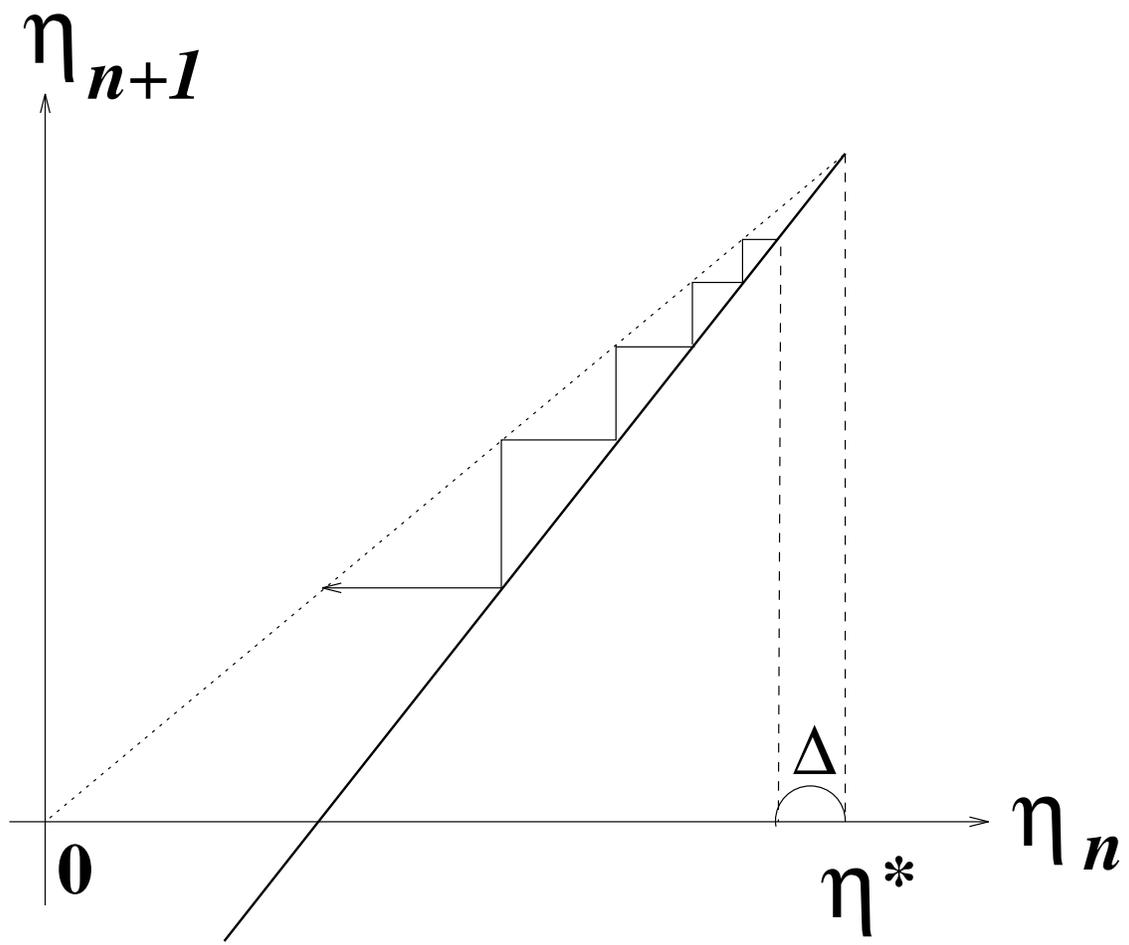
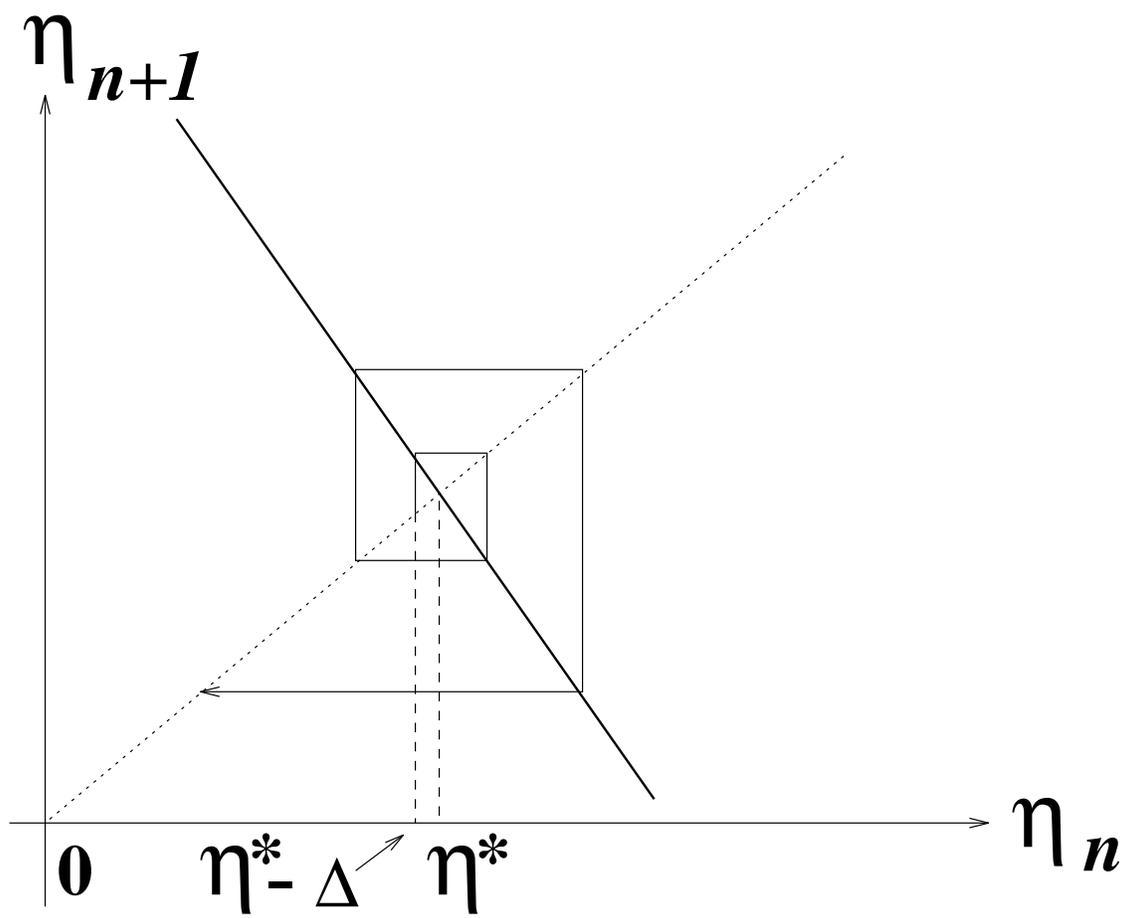
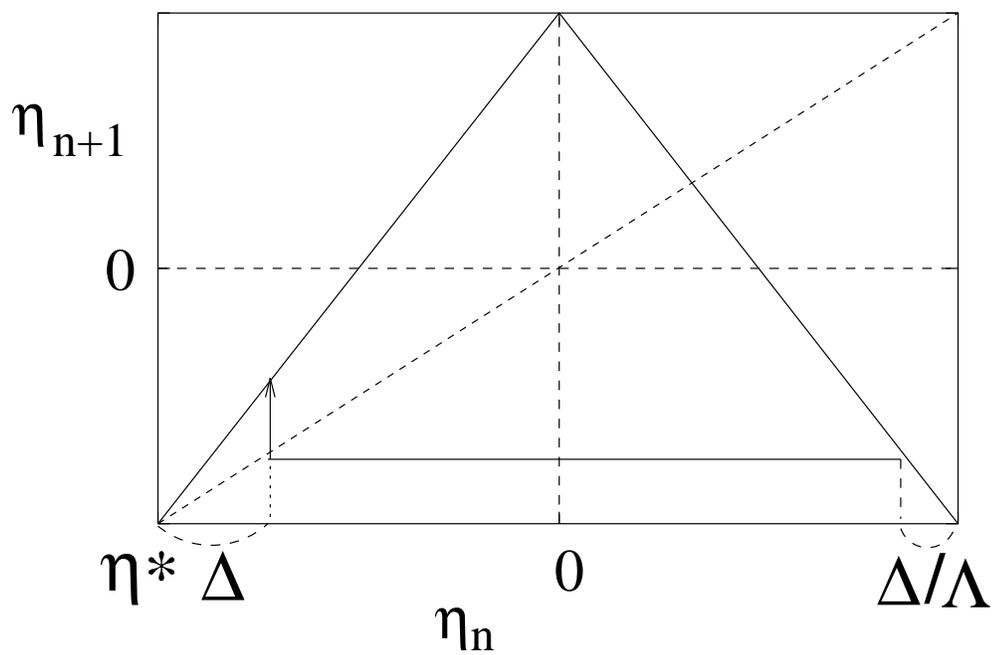
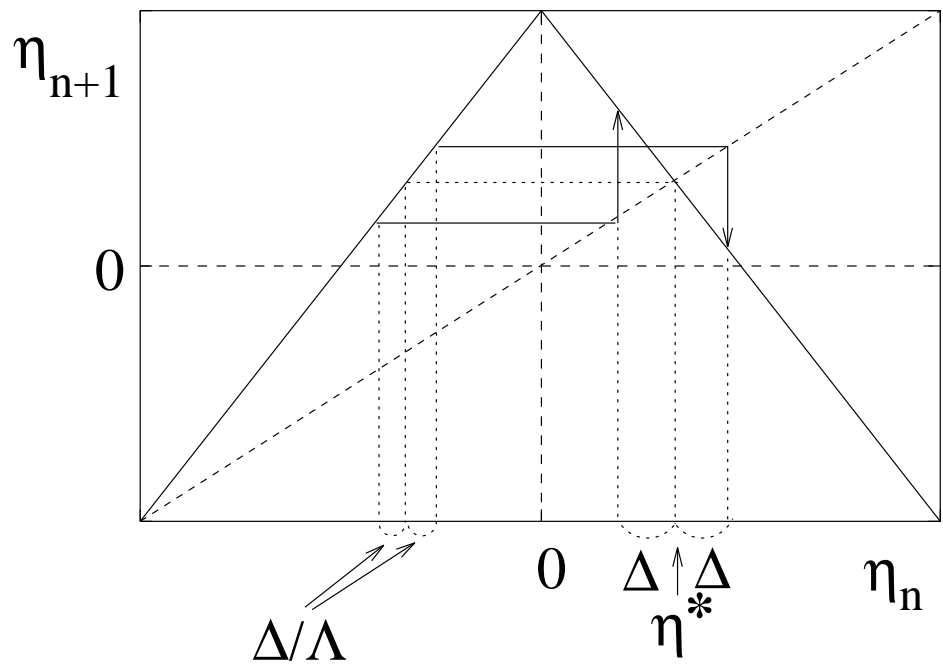
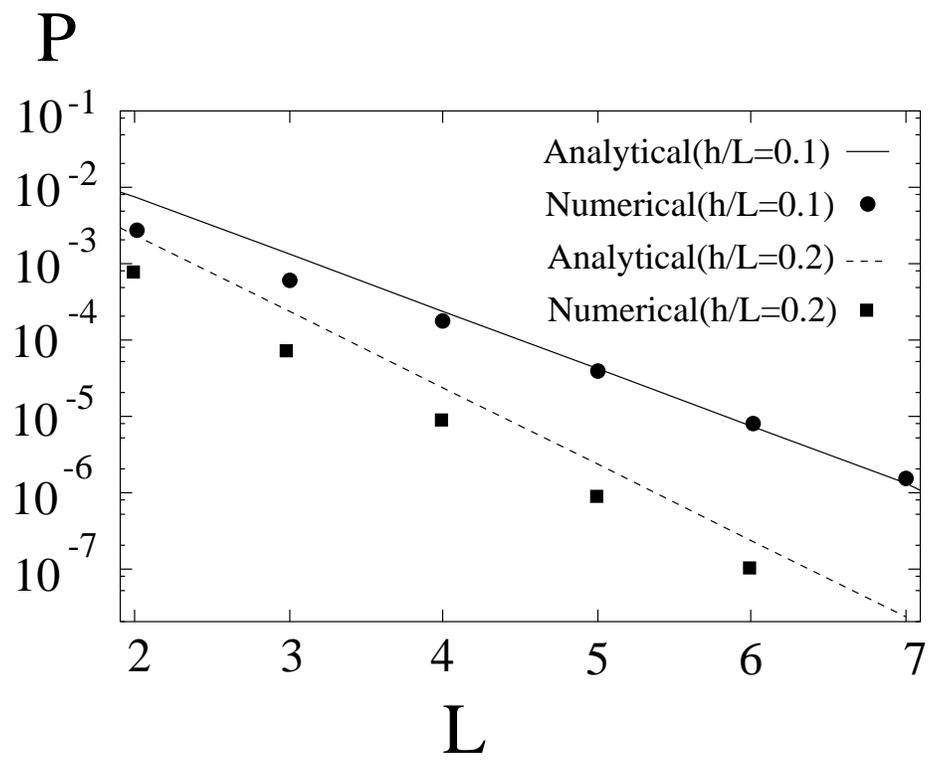
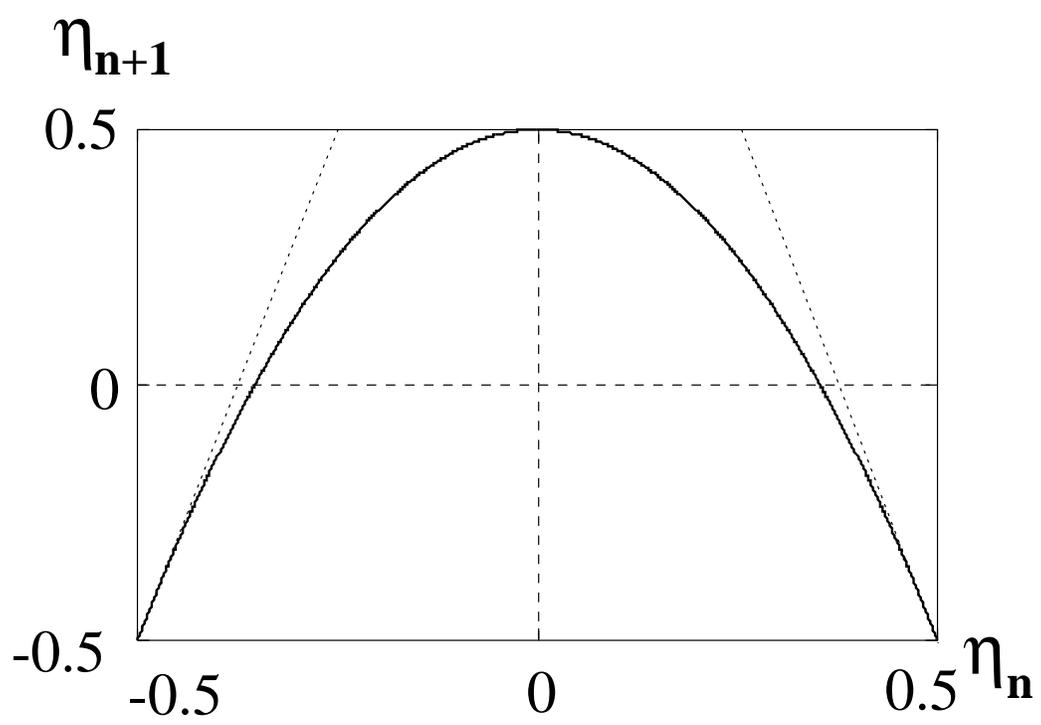
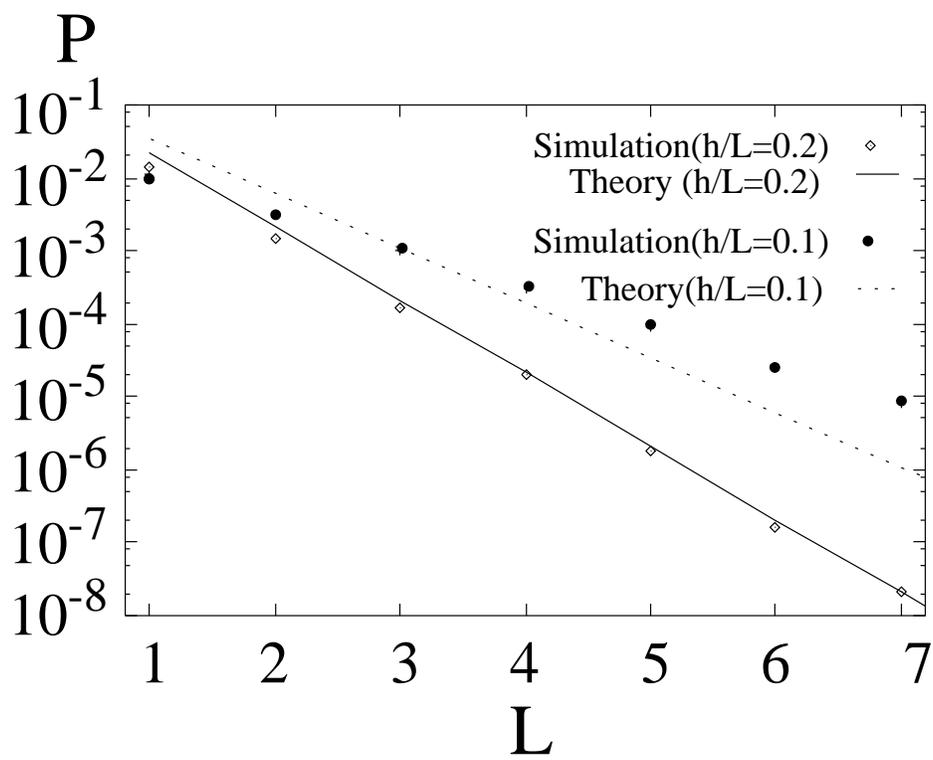
\end{document}